\documentclass[superscriptaddress,twocolumn,amsmath,amssymb,showpacs,preprintnumbers,aps]{revtex4-1}
\usepackage[dvips]{graphics}
\usepackage{amssymb}
\usepackage{amsmath}
\usepackage{latexsym}
\usepackage{epsfig}
\usepackage{latexsym}
\usepackage{dcolumn}
\usepackage{bm}
\usepackage{pifont}

\begin{document}
 \title{Relative velocity of sliding of microtubules by the action of Kinesin-5}
\author{Sthitadhi Roy{\footnote{Corresponding author(E-mail: sunroy@iitk.ac.in}}}
\affiliation{Department of Physics, Indian Institute of Technology,Kanpur 208016, India.}

\begin{abstract}
Kinesin-5, also known as Eg5 in vertebrates is a processive motor with 4 heads, which moves on
filamentous tracks called microtubules. The basic function of Kinesin-5 is to slide apart two 
anti-parallel microtubules by simultaneously walking on both the microtubules \cite{kapitein}. We develop an analytical expression for the 
steady-state relative velocity of this sliding in terms of the rates of attachments and detachments of motor heads with the 
ATPase sites on the microtubules. We first analyse the motion of one pair of motor heads on one microtubule and then couple it to the 
motion of the other pair of motor heads of the same motor on the second microtubule to get the relative velocity of sliding.
\end{abstract}

\maketitle

\section{Introduction}

The protein constituents of the cytoskeleton of eukaryotic cells can be divided broadly into 
the following three categories: (i){\it filamentous} proteins, (ii) {\it motor} proteins, and (iii){\it accessory} proteins. The three classes of filamentous proteins which form the main scaffolding of the cytoskeleton, are: (a){\it actin}, (b){\it  microtubule}, and (c){\it intermediate filaments}. The three superfamilies of motor proteins are: (i){\it myosin} superfamily, (ii){\it kinesin} superfamily, and (iii){\it dynein} superfamily. Microtubules serve as tracks for kinesins and dyneins.\cite{resource}.

The main cellular mission of the molecular motor Kinesin-5 is to cross-link anti-parallel microtubules and to slide them apart, thus playing a critical role during bipolar spindle formation \cite{kaseda}. Kinesin-5 is absolutely essential for the dynamic assembly and function of the mitotic and meiotic spindles \cite{roof}.

In this work, I have closely followed the paper by Peskin and Oster \cite{oster}.

\section{Model}
\begin{figure}[hp]
\begin{center}
\includegraphics[width=0.6\columnwidth]{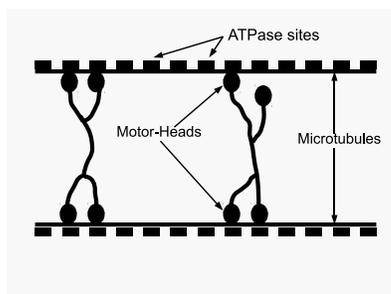}
\end{center}
\caption{Schematic diagram of Kinesin-5 on microtubules}
\label{schematic}
\end{figure}

\begin{figure}[hp]
\begin{center}
\includegraphics[width=0.8\columnwidth]{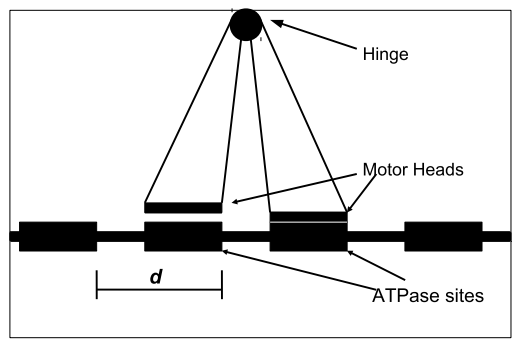}
\end{center}
\caption{Schematic diagram showing only one side of the motor with its two heads at discrete adjacent binding sites on a microtubule.}
\label{pair}
\end{figure}

The molecular motor Kinesin-5 is a four headed motor with two heads on each side. FIG-\ref{schematic} shows a schematic representation of the model we have developed for this motor.
In our model, we assume the two heads of the motor on one side to be joined at a hinge. Similarly, the other two heads on the other side are also joined by another hinge. These two hinges are then linked by a flexible connector. We also assume that the two heads on each side are free to rotate about their respective hinge. The two heads can also pass each other freely. However there is one restriction on their relative displacement. The maximum separation between two heads on one side can be equal to the separation between two discrete binding sites on the microtubule. This separation, which is a constant, is denoted by $ d $. We assume that the microtubules are aligned anti-parallel because Kinesin-5 has a much stronger preferences for cross-linking anti-parallel overlapped microtubules than parallel ones \cite{kaseda}. The $+$ end directed stepping activity of Kinesin-5 \cite{sawin} slides microtubules away from each other.  FIG.\ref{pair} shows schematically the model of one pair of heads on a microtubule.

The factors included in the model which are responsible for the motor to move forward are: (i) The back head has a higher probability to detach than the front head. (ii) The attached head prefers a geometry in which it leans in the forward direction i.e towards the $+$ end of the microtubule. The first of these factors ensures that the back head detaches and the second one ensures that it reattaches as a front head \cite{oster}. When both the heads are attached to the microtubules then the back head is in a more favourable orientation than the front one. Therefore, ATP hydrolysis and subsequent detachment is more likely at the back head. Once the back head detaches, the front head relieves itself of the strain by leaning forward and swinging the detached back head forward. Hence this is the process by which the motor moves forward.
The coordinates of the two heads are taken to be $x=x_1$ and $x=x_2$. The hinge is located halfway betwween them. The coordinate of the hinge, $x_h$ is therefore \[ x_h = \frac{x_1+x_2}{2} \] and the restriction dictates $|x_1-x_2|\le d$

Let

\begin{tabular}{lcl}
$\alpha$ &=& rate constant for attachment of the free head\\
&&to the track\\
$\beta_b$ &=& rate constant for detachment of the back head\\
&& from the track\\
$\beta_f$ &=& rate constant for detachment of the front head\\
&&from the track\\
$p$ &=& probability that the free head binds infront of\\
&&the bound head 
\end{tabular}
$\Rightarrow 1-p$ = probability that the free head binds behind the bound head.

Each head is also assigned a number which signifies its state. If it is attached to the track it is assigned $1$ and if it is detached from the track it is assigned $0$. 

\begin{figure}[hp]
\begin{center}
\includegraphics[width=0.8\columnwidth]{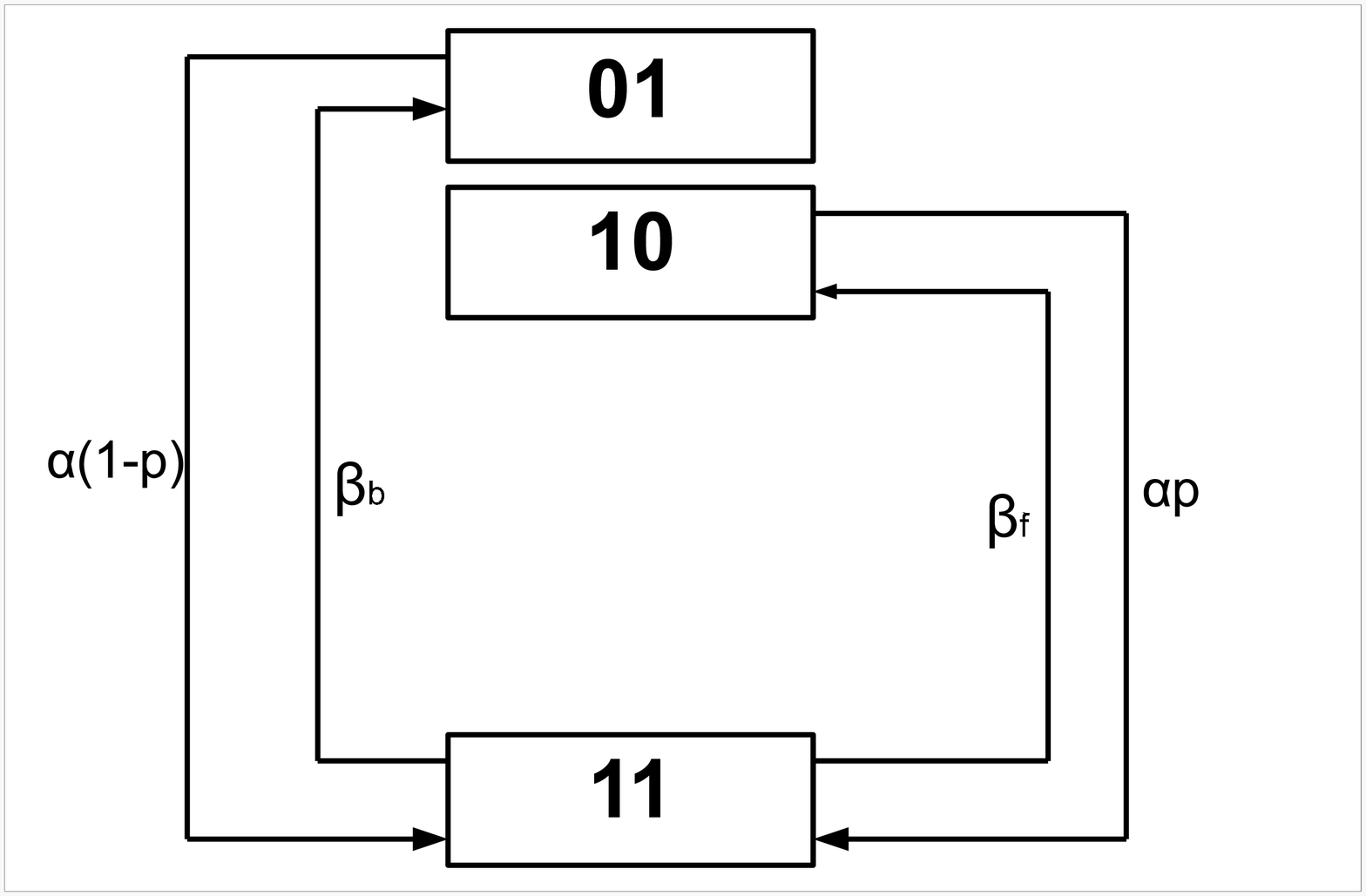}
\end{center}
\caption{Possible transitions in the states of one pair of heads of the motor and the transition rates}
\label{trans}
\end{figure}

The state $00$ corresponds to both the heads of the corresponding side completely detached from the track and that implies the termination of the walk. So we only consider the states $11$, $01$ and $10$. FIG.\ref{trans} shows the possible transitions and their rates.

In order to monitor the motion of one pair of heads on a microtubule and the velocity of this motion, it is simpler and adequate to monitor the motion of the mid-point of the heads \cite{oster}. The average velocity with which the mid-point moves on the microtubule can be taken as the average velocity with which the microtubule slides past the motor heads.
We define,

$x_m\equiv$ coordinate of the mid-point, which is identified as follows \cite{oster}:

$\Rightarrow x_m\equiv$ position of the hinge in the $11$ state

$\Rightarrow x_m\equiv$ position of he bound head in the $10\equiv01$ state 

\begin{figure}[hp]
\begin{center}
\includegraphics[width=0.8\columnwidth]{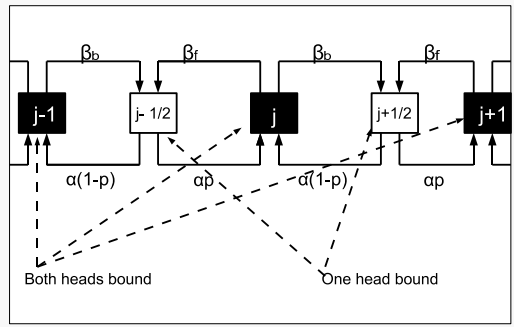}
\end{center}
\caption{The steps the mid-point can take and their respective rates}
\label{step}
\end{figure}

So when both the heads are attached then the mid-point is halfway between the two ATPase sites. As soon as one head detaches, the mid-point shifts to the location of the bound head i.e one of the two binding sites on either side of the previous location of the mid-point. So from FIG.\ref{pair} we can see that the midpoint jumps $\pm\frac{d}{2}$ and this is the step size of the motion of the mid-point. A state diagram showing the possible steps of the mid-point and their rates is shown in FIG.\ref{step}. 
It is assumed that the binding sites are labelled by half-integer indices and the mid-points between them are, consequently, labelled by integer indices. So when both the heads are attached, the mid-point is at an integer site and when only one head is attached, the mid-point is at a half-integer site.

\begin{figure}
\begin{center}
\includegraphics[width=0.8\columnwidth]{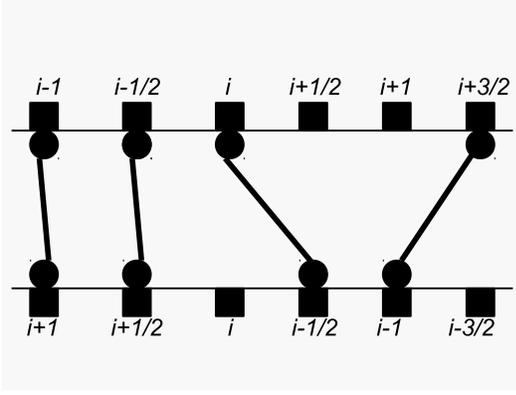}
\end{center}
\caption{The possible orientations of cross-links. The squares represent discrete sites where mid-points can be present. The circles represent the mid-points.}
\label{orientations} 
\end{figure}

So using the idea of Peskin and Oster explanined above, the motion of a pair of heads on one side can be simplified to the motion of the mid-point of the head alone. Similarly, the same thing can be done for the other pair of heads as well. So instead of monitoring the motion of four heads we can monitor the motion of the two mid-points like a two-headed motor. 
Now these two mid-points are coupled together because they are parts of the same 4-headed Kinsesin. This constraint introduces a restriction on the difference between the coordinates of the mid-points. If $x_{m1}$ and $x_{m2}$ are the coordinates of the mid-points then restriction is $|x_{m1}-x_{m2}|\le\frac{d}{2}$. So this ensures that if the mid-point is at a site $i$ on a microtubule then cross-linking can take place only through sites $i$ and $i\pm\frac{1}{2}$ on the other microtubule. Similarly, if the mid-point on a microtubule is at a site $i+\frac{1}{2}$, then the possible cross-linking sites on the other microtubule are $i+\frac{1}{2},i$ and $i+1$. The orientations allowed are shown in FIG.\ref{orientations}.

\section{Calculations and Results}

Let $P_j(t)\equiv$ probabilty of finding the mid-point at $j$ at time $t$. From FIG.\ref{step} we can write the following master equations:
\begin{equation}
\frac{dP_j(t)}{dt}={\alpha}pP_{j-\frac{1}{2}}+\alpha(1-p)P_{j+\frac{1}{2}}-(\beta_f+\beta_b)P_j
\label{diffint}
\end{equation}
\begin{equation}
\frac{dP_{j+\frac{1}{2}}(t)}{dt}=\beta_fP_{j+1}+\beta_bP_j-{\alpha}P_{j+\frac{1}{2}}
\label{diffhalfint}
\end{equation}
Now we define

$M_0=\sum^{\infty}_{j=-\infty}P_j$ and $N_0=\sum^{\infty}_{j=-\infty}P_{j+\frac{1}{2}}$.
\begin{eqnarray*}
\frac{dM_0}{dt}&=&\sum^{\infty}_{j=-\infty}\frac{dP_j}{dt}\\
&=&\sum^{\infty}_{j=-\infty}[{\alpha}pP_{j-\frac{1}{2}}+\alpha(1-p)P_{j+\frac{1}{2}}-(\beta_f+\beta_b)P_j]
\end{eqnarray*}
\begin{equation}
\Rightarrow\frac{dM_0}{dt}={\alpha}N_0-(\beta_f+\beta_b)M_0
\label{M0derivative}
\end{equation}
Similarly,
\[
	\frac{dN_0}{dt}=\sum^{\infty}_{j=-\infty}[\beta_fP_{j+1}+\beta_bP_j-{\alpha}P_{j+\frac{1}{2}}]
\]
\begin{equation}
\Rightarrow\frac{dN_0}{dt}=(\beta_b+\beta_f)M_0-{\alpha}N_0
\label{N0derivative}
\end{equation}
In the steady state $\frac{dM_0}{dt}=\frac{dN_0}{dt}=0$. So from equations (\ref{M0derivative}) and (\ref{N0derivative}) we get
\begin{equation}
{\alpha}N_0=(\beta_f+\beta_b)M_0
\label{ss1}
\end{equation}
Also from normalisation we have,
\begin{equation}
M_0+N_0=1
\label{ss2}
\end {equation}
Solving equations (\ref{ss1}) and (\ref{ss2}), we get
\begin{eqnarray}
M_0=\frac{\alpha}{\alpha+\beta_f+\beta_b} \label{M0}\\
N_0=\frac{\beta_f+\beta_b}{\alpha+\beta_f+\beta_b} \label{N0}
\end{eqnarray}

Mean position = ${\sum}xP(x)$

\[\Rightarrow \frac{Mean Position}{d}=\sum^{\infty}_{j=-\infty}[jP_j + (j+\frac{1}{2})P_{j+\frac{1}{2} }]\]
\[\Rightarrow \frac{Mean Speed}{d}=\frac{d}{dt}[\sum^{\infty}_{j=-\infty}[jP_j + (j+\frac{1}{2})P_{j+\frac{1}{2}}]]\]

If we define $<v>$ as the average speed of the mid-point then 
\[
\frac{<v>}{d}=[\sum^{\infty}_{j=-\infty}[j\frac{dP_j}{dt} + (j+\frac{1}{2})\frac{dP_{j+\frac{1}{2}}}{dt}].
\]
using the above relation and equations (\ref{diffint}) and (\ref{diffhalfint}) we can write
\begin{eqnarray*}
\frac{<v>}{d}&=&\sum^{\infty}_{j=-\infty} j[{\alpha}pP_{j-\frac{1}{2}}+{\alpha}(1-p)P_{j+\frac{1}{2}}-(\beta_f+\beta_b)P_j]\\
&&+\sum^{\infty}_{j=-\infty} \left(j+\frac{1}{2}\right)[\beta_bP_j+\beta_fP_{j+1}- {\alpha}P_{j+\frac{1}{2}}]
\end{eqnarray*}
Using $M_0=\sum^{\infty}_{j=-\infty}P_j$ and $N_0=\sum^{\infty}_{j=-\infty}P_{j+\frac{1}{2}}$, we get
\[\frac{<v>}{d}={\alpha}\left(p-\frac{1}{2}\right)N_0 + \left(\frac{\beta_b+\beta_f}{2}\right)M_0\].
Plugging in the expressions for $M_0$ and $N_0$ from equations (\ref{M0}) and (\ref{N0}), we get \cite{oster}
\begin{eqnarray}
\frac{<v>}{d}&=&{\alpha}\left(p-\frac{1}{2}\right)\left(\frac{\beta_f+\beta_b}{\alpha+\beta_f+\beta_b}\right) + \left(\frac{(\beta_b+\beta_f)\alpha}{2({\alpha+\beta_f+\beta_b})}\right) \nonumber \\
\Rightarrow<v>&=&\left(\frac{\alpha d}{\alpha+\gamma}\right)\left[\gamma\left(p-\frac{1}{2}\right)+\frac{\delta}{2}\right]\label{v}
\end{eqnarray}
where,
\begin{eqnarray*}
\gamma&=&\beta_b+\beta_f\\
\delta&=&\beta_b-\beta_f
\end{eqnarray*}
Now we couple the motion of the two mid-points which move on their MT tracks with $<v>$ calculated above. The motion of a mid-point counts only if it is attached to a microtubule and cross-linked to the other microtubule. So the relative velocity due to the motion of a mid-point on a microtubule is equal to $<v>$ multiplied to the probability of a mid-point being found at any particular site on that microtubule, which is then multiplied by the probability of a mid-point being found at all the corresponding possible cross-linking sites, and then summed over all the sites of the first microtubule.

This prescription can be mathematically expressed as, 

\begin{eqnarray}
<V>=&<v>_1&\sum^{\infty}_{i=-\infty}[P_i^{(1)}(P_i^{(2)}+P_{i+\frac{1}{2}}^{(2)}+P_{i-\frac{1}{2}}^{(2)})\nonumber\\
&& + P_{i+\frac{1}{2}}^{(1)}(P_i^{(2)}+P_{i+\frac{1}{2}}^{(2)}+P_{i+1}^{(2)})+\nonumber\\
&<v>_2&\sum^{\infty}_{i=-\infty}[P_i^{(2)}(P_i^{(1)}+P_{i+\frac{1}{2}}^{(1)}+P_{i-\frac{1}{2}}^{(1)})\nonumber\\
&& + P_{i+\frac{1}{2}}^{(2)}(P_i^{(1)}+P_{i+\frac{1}{2}}^{(1)}+P_{i+1}^{(1)})]\label{velocity}
\end{eqnarray}
Here $<v>_1$ and $<v>_2$ are the velocities of the mid-points on the first and second microtubules respectively. Similarly, the superscripts on the probabilities also refer to the the microtubules. But, since there is no functional difference between the two microtubules, these indices can be dropped. Then we have
\begin{equation}
<V>=2<v>[\sum_{i=-\infty}^{\infty}(P_i^2+P_{i+\frac{1}{2}}^2+4P_iP_{i+\frac{1}{2}}]
\label{velocity1}
\end{equation}
We define three new quantities,
\begin{eqnarray}
M&=&\sum_{i=-\infty}^{\infty}P_iP_{i+\frac{1}{2}} \label{M}\\
N&=&\sum_{i=-\infty}^{\infty}P_{i+\frac{1}{2}}^2 \label{N}\\
Q&=&\sum_{i=-\infty}^{\infty}P_i^2 \label{Q}
\end{eqnarray}
Plugging these expressions in equation (\ref{velocity1}),
\begin{equation}
<V>=2<v>(4M+N+Q)
\label{velocity2}
\end{equation}
Now,
\[\frac{dN}{dt}=2\sum_{i=-\infty}^{\infty}P_{i+\frac{1}{2}}\frac{dP_{i+\frac{1}{2}}}{dt}\] 
using equation (\ref{diffhalfint})
\[\frac{dN}{dt}=2\sum_{i=-\infty}^{\infty}[P_{i+\frac{1}{2}}(\beta_fP_{i+1}+\beta_bP_i-{\alpha}P_{i+\frac{1}{2}})]\]
using the definitions of equations (\ref{M}) and (\ref{N})
\begin{equation}
\frac{dN}{dt}=2[(\beta_f+\beta_b)M-\alpha N]
\label{Nderivative}
\end{equation}
Similarly,
\[\frac{dQ}{dt}=2\sum_{i=-\infty}^{\infty}P_i\frac{dP_i}{dt}\]
using equation (\ref{diffint})
\[\frac{dQ}{dt}=2\sum^{\infty}_{i=-\infty}[P_i({\alpha}pP_{i-\frac{1}{2}}+\alpha(1-p)P_{i+\frac{1}{2}}-(\beta_f+\beta_b)P_i)]\]
using the definitions of equations (\ref{M}) and (\ref{Q})
\begin{equation}
\frac{dQ}{dt}=2[\alpha M +\alpha(1-p)M-(\beta_f+\beta_b)Q]
\label{Qderivative}
\end{equation}
In the steady state $\frac{dN}{dt}=\frac{dQ}{dt}=0$, so we have,
\begin{eqnarray}
2[(\beta_f+\beta_b)M-\alpha N]&=&0 \label{final1}\\
2[\alpha M +\alpha(1-p)M-(\beta_f+\beta_b)Q]&=&0\label{final2}
\end{eqnarray}
Also since the variables $M$,$N$ and $Q$ capture all the cross-linking probabilities (FIG.\ref{orientations}) so we can write,
\begin{equation}
2M+N+Q=1
\label{final3}
\end{equation}
Solving equations (\ref{final1}), (\ref{final2}) and (\ref{final3}) we get,
\begin{equation}
M=\frac{\alpha(\beta_f+\beta_b)}{2\alpha(\beta_f+\beta_b)+(\beta_f+\beta_b)^2+\alpha^2(2-p)}
\label{Mvalue}
\end{equation} 
Using equations (\ref{final3}) and (\ref{Mvalue}), we get the final expression for $<V>$ as,

\begin{widetext}
\begin{eqnarray}
<V>&=&2<v>(1+2M)\label{finalvelocity}\\
i.e.,
<V>&=&2\left(\frac{\alpha d}{\alpha+\gamma}\right)\left[\gamma\left(p-\frac{1}{2}\right)+\frac{\delta}{2}\right]   \left(1+\frac{2\alpha(\beta_f+\beta_b)}{2\alpha(\beta_f+\beta_b)+(\beta_f+\beta_b)^2+\alpha^2(2-p)}\right)\label{finalvelocity2}
\end{eqnarray}
\end{widetext}

\section{Physical Interpretations}
From equation (\ref{Mvalue}) it is obvious that $M>0$. Therefore, $<V>\ge2<v>$ which is also consistent with our intuitive expectation. Since in our model cross-linking is allowed through sites next to the adjacent sites on microtubules, hence the two mid-points move with respect to each other. This factor actually increases the relative velocity above $2<v>$. The microtubules slide with respect to the mid-points with veocity $<v>$ and the mid-points move with respect to each other which increases the overall relative velocity.

Next we consider a few special cases to establish that our results are consitent with physical intuition.
\medskip

If the probability of the detachment of the back head and the front head is the same and the probability of the free head binding infront and behind the attached head is also the same, then
\begin{eqnarray}
\delta=&\beta_f-\beta_b=&0\label{sc1}\\
p&=\frac{1}{2}\label{sc2}
\end{eqnarray}
In that case, using equations (\ref{sc1}) and (\ref{sc2}) in equations (\ref{v}) and (\ref{finalvelocity2}), we get
$<v>=0$ and $<V>=0$ respectively, which is expected on physical grounds.

\medskip

If possible orientations for cross-linking are restricted by imposing the condition that cross-linking is allowed only through adjacent sites on two microtubules, then the relative positions of the mid-points with respect to each other do not change. The relative velocity of sliding will simply be the sum of the velocities with which each mid-point hops on the microtubule. The velocites are added because of the anti-parallel orientations of the microtubules. Mathematically, in this case 
\begin{equation}
M=\sum_{i=-\infty}^{\infty}P_iP_{i+\frac{1}{2}}=0
\label{sc3}
\end{equation}
and hence, using equation (\ref{sc3}) in equation (\ref{finalvelocity}) we get
\begin{equation}
<V>=2<v>,
\label{sc4}
\end{equation}
which agrees with the above prediction.
\medskip

\section{Comparison with earlier works}
Kruse et al.\cite{kruse} calculated the relative velocity of sliding between two filaments due to the action of a two headed motor. In this work the length of the filaments, which is denoted by $L$ is assumed to be finite. It is assumed that cross-linking can take place only through adjacent sites of the two microtubules. A variable $\xi$ is defined, which is the differnce between the coordinates of the minus ends of the microtubules. Thus overlaps can take palce if $\xi=0,...,2L-2$. The average velocity of sliding as a function of $\xi$ in units of monomers per unit time is
\begin{equation}
  v(\xi)=2p_{cl}
	\left\{
		\begin{array}{ll}
			\sum_{i=1}^{1+\xi}J_iS_{2+\xi-i} & \mbox{for} \xi=0,...,L-2,\\	
			\sum_{i=\xi-L+2}^{L-1}J_iS_{2+\xi-i} & \mbox{for} \xi=L-1,...,2L-3.
		\end{array}
	\right.
\label{kruse1}
\end{equation}
where $p_{cl}$ is the probability of cross-linking.

When averaged over all possible values of $\xi$ the final expression for average relative velocity is \cite{kruse}
\begin{equation}
<V>=\frac{1}{2L-1}\sum_{\xi=0}^{2L-3}v(\xi)
\label{kruse2}
\end{equation}
Now, equation (\ref{kruse2}) can be used for our system by taking the limit of $L\rightarrow\infty$ and replacing $v(\xi)$ by $2<v>$. In that case,
\begin{eqnarray*}
	<V>&=&\lim_{L\rightarrow\infty}\frac{1}{2L-1}\sum_{\xi=0}^{2L-3}v(\xi)\\
	   &=&\lim_{L\rightarrow\infty}\frac{1}{2L-1}\sum_{\xi=0}^{2L-3}2<v>\\
	   &=&\lim_{L\rightarrow\infty}\frac{2<v>}{2L-1}\sum_{\xi=0}^{2L-3}1\\
	   &=&2<v>\lim_{L\rightarrow\infty}\frac{2L-3}{2L-1}\\
	<V>&=&2<v>
\end{eqnarray*}
which is consistent with equation (\ref{finalvelocity2}) because in the special case cross-linking was allowed only through adjacent sites on two microtubules. 

Computer simulations as well as analytical studies related to continuum models of motor induced sliding of microtubules and actin filaments have been reported \cite{mogilner,krusecont}. These studies say that two headed bipolar motors slide anti-parallel filaments with a velocity that is twice that of the free motor velocity \cite{mogilner}. In our case, the motion of Kinesin-5 is reduced to a motion of a two headed motor using the model of mid-points. In our case cross-linking is allowed through adjacent sites on microtubules or sites next to adjacent sites on microtubules. In the continuum limit the distance between two sites is negligible, so it can be said that all the cross-linking that takes place is effectively through adjacent sites only, as in the continuum limit the adjacent site and the next to adjacent site can be deemed to be the same. Hence all the cross-linking can be deemed to be through adjacent sites. In such a case as we calculated in equation (\ref{sc4}) our realtive velocity of sliding $<V>$ is equal to $2<v>$, where $<v>$ is the velocity with which a mid-point slides a filament. If $<v>$ is taken analogous to the free motor velocity of a two headed motor, then our result in the continuum limit agrees with earlier studies done with continuum models.

\section{Acknowledgements}
I sincerely thank Debashish Chowdhury for introducing me to molecular motors, and for useful suggestions. I also thank him for his comments on earlier versions of this manuscript.

\end{document}